\documentclass[conference]{IEEEtran}
\IEEEoverridecommandlockouts
\usepackage{cite}
\usepackage{amsmath,amssymb,amsfonts}
\newtheorem{definition}{Definition}
\usepackage{algorithmic}
\usepackage{graphicx}
\usepackage{textcomp}
\usepackage{xcolor}
\usepackage{seqsplit}
\usepackage{tikz}
\usepackage{tabularx}
\usepackage{array}
\usepackage{listings}
\usepackage{adjustbox}
\usepackage{multirow}
\newcolumntype{M}[1]{>{\centering\arraybackslash}m{#1}}
\usetikzlibrary{shapes.geometric,shapes.symbols,fit,positioning,shadows, arrows.meta, backgrounds}

\def\BibTeX{{\rm B\kern-.05em{\sc i\kern-.025em b}\kern-.08em
    T\kern-.1667em\lower.7ex\hbox{E}\kern-.125emX}}
\begin{document}
 \makeatletter
\newcommand{\linebreakand}{%
\end{@IEEEauthorhalign}
\hfill\mbox{}\par
\mbox{}\hfill\begin{@IEEEauthorhalign}
}
\makeatother
\title{A Constraint-based Recommender System via RDF Knowledge Graphs 
}
\author{
	\IEEEauthorblockN{Ngoc Luyen Le\IEEEauthorrefmark{1}\IEEEauthorrefmark{2}\\\textit{ngoc-luyen.le@hds.utc.fr}}
	\and
	\IEEEauthorblockN{Marie-Hélène Abel\IEEEauthorrefmark{1}\\\textit{marie-helene.abel@hds.utc.fr}}
	\and
	\IEEEauthorblockN{Philippe Gouspillou\IEEEauthorrefmark{2} \\\textit{p.gouspillou@vivocaz.fr} }
	\linebreakand
	\IEEEauthorblockA{
	\IEEEauthorrefmark{1}Université de technologie de Compiègne, CNRS, Heudiasyc (Heuristics and Diagnosis of Complex Systems),\\ CS 60319 - 60203 Compiègne Cedex, France.
	\\
	\IEEEauthorrefmark{2}Vivocaz, 8 B Rue de la Gare, 02200, Mercin-et-Vaux, France.
}
}


\maketitle

\begin{abstract}

Knowledge graphs, represented in RDF, are able to model entities and their relations by means of ontologies. The use of knowledge graphs for information modeling has attracted interest in recent years. In recommender systems, items and users can be mapped and integrated into the knowledge graph, which can represent more links and relationships between users and items. Constraint-based recommender systems are based on the idea of explicitly exploiting deep recommendation knowledge through constraints to identify relevant recommendations. When combined with knowledge graphs, a constraint-based recommender system gains several benefits in terms of constraint sets.
In this paper, we investigate and propose the construction of a constraint-based recommender system via RDF knowledge graphs applied to the vehicle purchase/sale domain. The results of our experiments show that the proposed approach is able to efficiently identify recommendations in accordance with user preferences.

\end{abstract}

\begin{IEEEkeywords}
Knowledge graph, Constraint-based Recommender System, Ontology
\end{IEEEkeywords}

\section{Introduction}

The right item recommendation at the right moment for a user is always the goal pursued by any recommender system. With the ever-growing volume of information in various applications, recommender systems are a useful way to overcome information overload and allow users to explore new opportunities and suggestions in a personalized way by matching their preferences. Therefore, the relevant recommendations of a recommender system increasingly influence users' decisions in choosing a service, product, or content as well as enhancing user experiences on the platform.

In several domains such as financial services, expensive luxury goods, real estate, or automobiles, these items are purchased less frequently and are commonly more costly than convenience others. Therefore, finding item recommendations requires the high involvement of users by providing their preferences or needs. In other words, the recommender system attempts to retrieve relevant recommendation items from the user's answers to a set of questions about their preferences for items. Therefore, constraint-based recommender systems are a typical approach in these complex domains. 

In constraint-based recommender systems, the identification of recommendations is considered a process of constraint satisfaction. Therefore, constraints play an extremely important role in this type of recommender system. Some constraints may come from the item domain that includes domain-specific knowledge about items. Other constraints may come from users that rely on their preferences about items \cite{felfernig2008constraint}.  The combination of two types of constraints causes an increase in the search space of an item. Besides, this may lead to the repetition of the same type of constraints for groups of users who shared some common characteristics. The use of RDF knowledge graphs with the support of ontologies can help to reduce the set of constraints by using reasoning mechanisms to deduct relevant information about domain-specific knowledge. Therefore, we propose in this paper our approach for the construction of a constraint-based recommender system via RDF knowledge graphs.



The remainder of this paper is organized as follows. In the following section, we introduce works from the literature on RDF knowledge graphs and constraint-based recommender systems. Section III presents our main contributions to the construction of the constraint-based recommender system exploiting RDF knowledge graphs. In section IV, we experiment with our proposed approach based on an RDF knowledge graph in the vehicle purchase/sale domain. Finally, we conclude the paper with some ideas for future work in the last section.
\section{Related Works}
\subsection{Knowledge representation by means of Ontology}

Knowledge representation focuses on studying the representation form for knowledge and how it is computed and used within machines. More specifically, knowledge representation concerns with capturing and presenting information in a form that a machine can understand and utilize to solve complex problems \cite{stephan2007knowledge}. In the context of knowledge sharing, ontology is used as a knowledge representation for knowledge bases. In general, an ontology is a formal and explicit description of shared knowledge that consists of a set of concepts in a domain and the relationships between those concepts \cite{guarino1995towards}.

An ontology plays a role as the backbone of the formal semantics of a knowledge graph. Basically, ontologies can be expressed in Resource Description Framework (RDF) Schema and Web Ontology Language (OWL) by a set of RDF triples. An RDF triple is defined as a set of three components: a subject, a predicate, and an object. A triple $\langle subject, predicate, object \rangle$ expresses that a given subject has a given value for a given property \cite{mcguinness2004owl}. Intuitively, if the subject and the object are two nodes in a graph, the predicate describes the relationship between these two nodes.  An ontology represented in OWL owns a reasoning mechanism that allows the deduction of additional knowledge.

The construction of an ontology from a domain can be done through several methods, including manual and automated approaches\cite{ffrez_2002}. In a manual approach, domain experts are typically involved in the ontology development process to ensure that the ontology is consistent with the domain knowledge. However, expert dependency is not always necessary, as ontologies can also be developed based on existing resources, such as taxonomies, dictionaries, and databases. Automated approaches involve natural language processing, machine learning, and other techniques to extract concepts, relationships, and entities from unstructured or semi-structured data.


While an RDF knowledge graph can be considered as a type of ontology, not all RDF knowledge graphs are ontologies, and the primary focus of an RDF knowledge graph is on representing data and relationships between entities, rather than defining a formal vocabulary for a specific domain. In the next section, we will provide more details on recommender systems and specify the role that an RDF knowledge graph can play, particularly in the context of constraint-based recommender systems.
\subsection{Recommender Systems}
Recommender systems are a special application that estimates users' preference for items and attempts to recommend the most relevant items to users through information retrieval. The suggestions provided by a recommender system help to support users in various decision-making processes such as what music to listen or what products to purchase. In general, recommender systems are usually classified into main six categories: Collaborative Filtering RSs, Content-based RSs, Demographic-based RSs, Knowledge-based RSs, Context-aware RSs, and Hybrid RSs  \cite{le2022towards}. 

If the amount of collected data is limited, the results of systems such as Collaborative Filtering RSs, Content-based RSs, and Demographic-based RSs can either be poor or lack full coverage over the spectrum of combinations between users and items. Indeed, these approaches can face some problems such as cold start, data sparsity, limited context analysis, and over-specialization \cite{adomavicius2005toward, ramezani2008selecting}. Knowledge-based recommender systems are proposed to tackle these problems based on explicitly soliciting user preferences for such items and deep knowledge about the domain to compute relevant recommendations \cite{felfernig2011developing}. In particular, this type of recommender system is well-suited for situations where (i) users wish to specify their requirements explicitly; (ii) it is difficult to achieve feedback for items; and (iii) feedback may be out-of-date or time-sensitive. For example, if an item is a used car, the feedback may not be very useful for computing recommendations because a used car is purchased only once.

By considering the way of user interactions and the corresponding knowledge base used for these interactions, there are two types of knowledge-based RSs: namely constraint-based recommender systems \cite{felfernig2015constraint} and case-based recommender systems \cite{bridge2005case}. While case-based recommender systems find similar items by computing and adapting recommendations based on similar cases in the past. In constraint-based recommender systems, a set of rules/constraints will be defined to match user preferences/user requirements to item properties. 
Constraint-based recommender systems have been applied in different domains to help users adopt the best relevant item recommendations. In \cite{boudaa2021datatourist, jannach2009constraint}, the authors developed constraint-based recommender systems based on the usage of knowledge bases in the tourism domain. In \cite{atas2019towards}, the author proposed the amelioration of using of constraint-based recommender system by similarity over user requirements. Using rules/constraints become more popular to improve recommendation results such as in e-commercial application \cite{dadouchi2022context}, in simulation systems \cite{le2023constraint}, or in financial services \cite{felfernig2016application}.

The purchase and sale of used vehicles are not as frequent as other products, and each vehicle item has only one transaction. In general, user preferences for their favorite vehicles play an important role in recommending relevant used vehicles. Therefore, we propose constructing a constraint-based recommender system for vehicle purchases/sales using RDF knowledge graphs. In the next section, we will present our proposal for this work in detail.

\section{Our proposition}
In this section, we present our proposal for constructing a constraint-based recommender system using an RDF knowledge graph. To illustrate our approach, we use ontologies from the e-commerce domain related to the purchase and sale of vehicles to create a knowledge base.
\subsection{RDF Knowlege Graph}
In the context of an e-commercial application, the construction of a knowledge base for the vehicle domain consists of three main axes: item-vehicle attributes, user-buyer profiles, and interactions between user-buyer and item-vehicle. 
The collection of this information can be organized and rewritten as triples, formally defined as 
$G_V$ = $\{a_{1}^{v},$ $a_{2}^{v},$ $...,$ $a_{n}^{v}$$\}$
where $a_{i}^{v}$ presents a complete RDF triple $a_{i}^{v} = \langle subject_i, predicate_i, object_i\rangle$. Similarly, user profiles that include user information and user preferences about vehicles can be also defined as a set of RDF triples: $G_U$ $=$ $\{a_{1}^{u},$ $a_{2}^{u},$ $...,$ $a_{m}^{u},\}$
where $a_{j}^{v}$ presents a complete RDF triple. Finally, when a user adds an item to their list of favorite items, it means that this item is interesting to the user. These interactions between users and items are defined as: $RS$ $:$ $G_U$ $\times$ $G_V$ $\times$ $G_C$ $\longrightarrow$ $Interaction$
where $G_U$ corresponds to the user, $G_V$ denotes the vehicle description item, and $G_C$ expresses contextual information concerning the user and the item when the interaction is done, for example, user objectives, times, locations, and resource information. In our work, we use the ontology developed for vehicle description and user profile that we presented in \cite{le2022towards}. 

In this paper, we organize user profiles and vehicle descriptions under an RDF knowledge graph. Figure \ref{fig_01} illustrates an example of RDF knowledge graphs. This graph is equivalent to a set of triples. Each edge of the graph is demonstrated as the predicate of the triple, the source node is represented as the subject of the triple, and the destination node is described by the object of the triple.
\begin{figure}[h!t]
	\vspace{-0.3cm}
	\centering
	\includegraphics[width=0.40\textwidth]{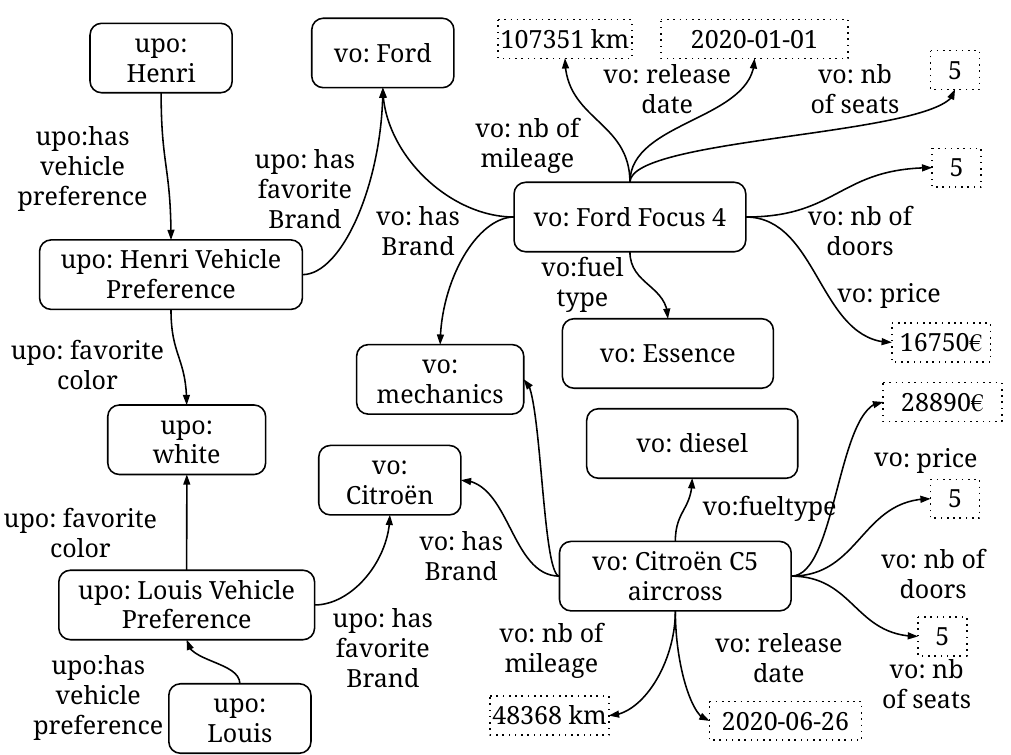}
	\caption{An extracted RDF graph representation of user preferences and vehicle descriptions}
	\label{fig_01}
	\vspace{-0.3cm}
\end{figure}

The structure and organization of user and item information under the sets of RDF triples allow an explicit representation of user profiles and vehicle descriptions. This type of representation supports not only a semantic search capacity instead of searching by keywords but also the capacity of integration of data from different sources \cite{enellart_2011}. Furthermore, RDF knowledge graphs can assist in reasoning relevant and insightful information based on ontologies. 

\subsection{A Constraint-based Recommender System}
Having defined the basics of an RDF knowledge graph for the purchase/sale of vehicles, we show in this section how to define and construct a constraint-based recommender system based on this data source.

User requirements can be gathered through different forms such as user query history, user contextual information, or user responses from the set of questions on their preference. In our work, we focus on treating user requirements from user preferences and their contextual information. Firstly,  the preferences of users about their favorite vehicle are considered as a part of the information in user profiles. Therefore, users need to provide their preferences related to the characteristics of the vehicle that they would like to own. For example, several users may have a preference for \textit{black} or \textit{white color} for their vehicles, or other users want a \textit{vehicle with 7 places for the family}. Second, the contextual information of the user can be the external situations. For example, the location where users live or work might be an important factor in selecting vehicle types. Therefore, information on user preferences and user context play a role as constraints in order to filter relevant recommendation items for users. 

On the other side, we figure out the bridge between user requirements and vehicle description items by using the vehicle descriptions and domain knowledge. Firstly, the vehicle description encompasses the properties of a given item, while domain knowledge provides deeper insights about the items. For instance, when a user declares their profile and expresses interest in a ``\textit{family profile}'', the domain knowledge for vehicle items enables the recommendation of large-size vehicles that have \textit{a number of places} greater than \textit{three seats}.

Constraint-based recommendation relies on the exploration of relations between user requirements and item properties. The knowledge base in our case can be considered as a set of variables and a set of constraints. Using these variables and constraints can constitute the elements of a constraint satisfaction problem (CSP) \cite{felfernig2006integrated, felfernig2011developing}. The solutions of this CSP allow for finding the most relevant recommendations in a recommender system. The task of calculating and suggesting recommendations for a user from their preference is called a recommendation task. A constraint-based recommender system is able to compute and propose recommendations from a recommendation task.

\begin{definition} A recommendation task  is defined as a CSP($\mathcal{V}_U$,$\mathcal{V}_I$,$\mathcal{C}$), where $\mathcal{V}_U$ =$\{vu_1, vu_2, ..., vu_n\}$ denotes a set of variables that represent user preferences, $\mathcal{V}_I$ = $\{vi_1, vi_2, ..., vi_m\}$ is a set of  variables that represent item properties, $\mathcal{C} = \mathcal{C}_{KB} \cup \mathcal{C}_F$ refers to the set of constraints representing domain-specific constraints $\mathcal{C}_{KB}$ and the set of filter constraints $\mathcal{C}_F$ that describe the link between user preferences and items.
\label{def1}
\end{definition}

In the context of an e-commerce application for the purchases/sales of vehicles, we can extract different user preferences as a set of variables for $\mathcal{U}$ and vehicle item properties as the set of variables for $\mathcal{V}$. In particular, we illustrate the sets of variables by a simple example as follows:
\begin{itemize}
	\item  $\mathcal{V}_U$ = $\{vu_1: vehicleType(sedan, crossover, suv)$,\newline $vu_2:color(blue, black, white, red)$,\newline
	 $vu_3:profile(studentProfile, parentProfile, \newline businessProfile, professionalProfile)$,
	 $vu_4:seats(integer)$,  $vu_5:maxMileage(integer)$, \newline $vu_6:brand(text)$, $vu_{7}:maxBudget(integer)\}$
	\item  $\mathcal{V}_I$ = $\{vi_1: name (text)$,\newline $vi_2:price(integer)$, $vi_3:bodyType(text)$,
	\newline $vi_4:seats(2-places, 4-places, 5-places)$, 	\newline $vi_5:modelYear(2021,2020, 2019,2018)$, 
	\newline $vi_6:brand(Peugeot, Renault, Citroen, Tesla)$, 
	\newline $vi_7:mileage(integer)\}$
\end{itemize}
Every constraint can be classified into $\mathcal{C}_{KB}$ or $\mathcal{C}_F$. While the $\mathcal{C}_{KB}$ constraints are formed from using knowledge of the domain. $\mathcal{C}_F$ define particular requirements of the user on items. We can show several examples of the $\mathcal{C}_{KB}$ and $\mathcal{C}_F$ constraints in  Table \ref{table_01} and \ref{table_02}.

\begin{table}[h]
	\begin{center}
		\begin{tabular}{|p{0.7cm} | p{6cm} |}
			\hline 
			\textbf{Name} & \textbf{Constraint description} \\\hline
			$\mathcal{C}_{KB1}$ & The technical inspection dating from less than 6 months is required for a used vehicle more than 4 years old.   \\\hline
			$\mathcal{C}_{KB2}$ & If users prefer long distance routes, SUV or Crossover may suit them.    \\\hline
		\end{tabular}
	\end{center}
	\caption{Example of domain-specific knowledge constraints} 
	\label{table_01}
	\vspace{-0.3cm}
\end{table}

\begin{table}[h]
	\begin{center}
	\begin{tabular}{|p{0.7cm} | p{6cm} |}
		\hline 
		\textbf{Name} & \textbf{Constraint description} \\\hline
		$\mathcal{C}_{F1}$ & the price of the item has to be lower than or equal to the maximum budget of the user. \\\hline
		$\mathcal{C}_{F2}$ & the number of mileage of the item must be lower than the maximum mileage imposed by the user.  \\\hline
		$\mathcal{C}_{F3}$ & the number of place of the item has to equal to the number seats required by the user.  \\\hline
		$\mathcal{C}_{F4}$ & the color of the item has to be either white or blue.  \\\hline
	\end{tabular}
	\end{center}
\caption{Example of constraints related to user preferences}
\label{table_02}
\vspace{-0.5cm}
\end{table}

\begin{definition} A recommendation (a solution) for a given recommendation task ($\mathcal{V}_U$,$\mathcal{V}_I$,$\mathcal{C}$) is defined as an instantiation of $\mathcal{V}_I$ by realizing a complete assignment to the variables of ($\mathcal{V}_U$,$\mathcal{V}_I$) such that the constraints in $\mathcal{C}$ are satisfied. The recommendation is \textit{consistent} if the assignments are \textit{consistent} with the constraints.
\label{def2}
\end{definition}
Constraint-based recommender systems rely on an explicit knowledge base of the domain of users and items. With two types of constraints, we can compute relevant recommendations for a user. The constraints of $\mathcal{C}_{KB}$ related domain-specific knowledge can be resolved by using rules which are integrated into ontologies. Therefore, we will explore this approach in the next section based on the ontology model of the authors in \cite{le2022towards, le2022apport} and the RDF knowledge graph for user profiles and vehicle descriptions.

\subsection{Domain-specific knowledge constraints by SWRL rules}

In the context of the vehicle purchase/sale domain, ontologies are used to structure and organize the descriptions of vehicles and user profiles. The proposed ontology \cite{le2022towards} is constructed using OWL, which is a highly expressive, flexible, and efficient knowledge representation language based on the mathematical background of Description Logic. OWL can realize the reasoning of implicit information by processing explicit knowledge, which improves information management. Reasoning on ontologies results in a reduction in time, effort, and performance on ontologies \cite{miled2020knowledge}. Rules are useful to implement the deductive part of the knowledge base. In our work, we use the Semantic Web Rule Language (SWRL) to write rules on RDF knowledge graphs.

Formally, a SWRL rule consists of a high-level abstract syntax that contains a condition part and a conclusion part. The constitution of the condition and conclusion parts is positive conjunctions of atoms. The meaning of a SWRL rule can be explained as follows: if all the atoms in the condition are satisfied, then the conclusion must occur \cite{horrocks2004swrl}. In various applications, rules are necessary to extend the expressiveness of OWL. Therefore, using rules in conjunction with ontologies in such cases becomes an efficient way to solve problems \cite{golbreich2005reasoning}.

The constraints in the set of constraints $\mathcal{C}_{KB}$ often apply to a class, properties of a class, or group of individuals. In other words, these constraints affect global information within the scope of the knowledge base. These constraints can be translated into rules to be integrated into the ontology using SWRL. For example, for the constraint $\mathcal{C}_{KB2}$ that is used for all users who have a preference for the \textit{long distance route}, we can employ a SWRL rule to deduce the \textit{vehicle type} preferred by the user. Therefore, we propose to represent domain-specific knowledge constraints using SWRL rules, based on the advantages in information deduction.

\begin{table}[h]
	\begin{center}
		\begin{adjustbox}{width=.42\textwidth,center}
	\begin{tabular}{|m{2.8em} | m{25.5em}|}
		\hline
		\textbf{Name} & \textbf{SWRL Rule Expression} \\\hline
		$\mathcal{C}_{KB1}$ &  
		$Automobile(?a)$ $\wedge$ $Check(?c)$ $\wedge$ $inspected(?a,$ $?c)$   $\wedge$ $productionDate(?a,$ $?pdate)$ $\wedge$ $validFrom(?c,$ $?cdate)$ $\wedge$ $temporal:duration$$(?pduration,$ $?pdate,$ $``now",$ $``Months")$ $\wedge$ $temporal:duration$$(?cduration,$ $?cdate,$ $``now",$ $``Months")$ $\wedge$ $swrlb:greaterThan$$(?pduration,$ $48)$ $\wedge$ $swrlb:greaterThan$$(?cduration,$ $6)$ $\longrightarrow$ $isRequired$$(?c,$ $true)$
		\\\hline
		$\mathcal{C}_{KB2}$ &  $VehiclePreference(?vpu)$ $\wedge$ $hasFavoriteRouteType(?vpu,$ $?route)$ $\wedge$ $sameAs(?route,$ $upo:longDistanceRoute)$ $\longrightarrow$ $hasFavoriteVehicleType(?vpu,$ $upo:SUV)$ $\wedge$ $hasFavoriteVehicleType(?vpu,$ $upo:Crossover)$\\
		\hline
	\end{tabular}
\end{adjustbox}
	\end{center}
	\caption{The SWRL rules for the constraints  defined in the table \ref{table_01}}
	\label{tab:rule_tb1}
	\vspace{-0.5cm}
\end{table}

The SWRL rules provide powerful deductive capabilities based on integration with ontologies. However, SWRL is essentially a rule language, and it does not provide strong support for filtering and querying information from the RDF knowledge graph. Therefore, we will present an approach for constraints $\mathcal{C}_{F}$ related to user preferences that involves filtering and matching on RDF knowledge graphs in the next section.

\subsection{User preference constraints by SPARQL Queries}
SPARQL is a standard graph-matching query language designed to retrieve and manipulate data stored in RDF knowledge graphs on triplestores. A SPARQL query typically consists of three parts: (i) the pattern matching part which defines patterns used for variable matching such as optional, the union of pattern, nesting, or filtering declarations; (ii) the solution modifier part which adjusts outputs by modifying values of variables by several operators such as distinct, order, limit, or offset; and (iii) the output of the query defines a set of variables which match the patterns to return, or constructs new triples/graphs \cite{perez2009semantics}.

Suppose $Q$ is a SPARQL query and $c$ is a constraint. $Q$ FILTER $c$ is called a constraint query, where every variable in the constraint is satisfied in query $Q$. A solution of a SPARQL query $Q$ is defined as an assignment of variables in $Q$ to values. A set of possible values that can be assigned to a variable is called a domain. A recommendation or solution is consistent if all variables declared in the query have a corresponding value guaranteed. To find all possible solutions, we select a value in the RDF knowledge graph for each variable and match it with conditions from patterns and filters. In view of these aspects, finding recommendations for the constraint-based recommender system defined in definitions \ref{def1} and \ref{def2} is considered finding the solutions of a SPARQL query $Q$ with a set of constraints $c$. The equivalent expressions of these are described as follows:

\begin{itemize}
	\item The variables in $\mathcal{V}_U$ and $\mathcal{V}_I$ are used as the primary variables in the SPARQL query $Q$ on the RDF knowledge graphs associated with $G_U$ and $G_I$.
	\item Constraints $c \in \mathcal{C}_F$ must be satisfied by incorporating the FILTER clause into the SPARQL query $Q$.
\end{itemize} 
\begin{figure}[h]
\begin{adjustbox}{width=.35\textwidth,center}
\begin{lstlisting}[language=SPARQL, label=lst:sparql,basicstyle=\scriptsize\ttfamily,frame=lines, numbers=left,firstnumber=1,xleftmargin=2.5em,framexleftmargin=2.5em]
PREFIX uvso: <http://utc.fr/uvso/ns#> 
PREFIX uvo: <http://utc.fr/uvo/ns#> 
PREFIX uvoo: <http://utc.fr/uvoo/ns#>
PREFIX rdf: <http://w3.org/1999/02/22-rdf-syntax-ns#> 
PREFIX xsd: <http://w3.org/2001/XMLSchema#> 
PREFIX gr: <http://purl.org/goodrelations/v1#> 

SELECT ?auto
WHERE {
 ?auto rdf:type uvso:Automobile.
 ?auto uvso:color ?color. 
    FILTER contains(?color, "noir").
 ?auto uvso:seatingCapacity ?seats. 
 ?seats gr:hasValueInt "5"^^xsd:int.
 ?auto uvso:hasManufacturer ?brand. 
    FILTER (contains(str(?brand), "audi")).
 ?auto uvso:bodyStyle uvso:berline_occasion.
 ?auto uvso:mileageFromOdometer ?mileage .
 ?mileage gr:hasValueFloat ?mileageValue.
    FILTER (?mileageValue <= 100000) .
 ?auto uvo:valuation ?evaluation. 
 ?evaluation uvoo:hasCurrencyValue ?price. 
    FILTER (?price <= 100000 && ?price >= 20000) .
} LIMIT 10
\end{lstlisting}
\end{adjustbox}
	\caption{A SPARQL query by matching with a user preference}
	\label{fig_02}
	\vspace{-0.2cm}
\end{figure}

Graph pattern matching basically is the mechanism used by SPARQL in order to retrieve information from RDF knowledge graphs. A constraint in this context is considered an evaluation of a graph pattern on the RDF knowledge graph. To find solutions, SPARQL queries can use triple patterns and solution modifiers as constraints. Triple patterns involve three variables, and solution modifiers, such as ORDER BY, DISTINCT, and LIMIT, can be used to sort, eliminate duplicates, and limit solutions. This approach benefits from the expressivity of SPARQL queries, which have the expressive power of relational algebra.

\section{Experiments}
In order to evaluate our proposed approach, we use the RDF knowledge graph consisting of 5537 individuals of vehicle descriptions and 367 user preferences, which contain a total of 822,000 RDF triples based on ontological models presented in \cite{le2022towards}. Based on an empirical study of the RDF datasets, we demonstrate how our constraint-based recommender system via the RDF knowledge graph works. First, organizing user preferences and vehicle descriptions into RDF triples based on the ontology model enables us to gather data formally. The construction of a constraint-based recommender system then focuses on resolving two sets of constraints: domain-specific knowledge constraints and user preference constraints. In particular, the set of constraints based on domain-specific knowledge is translated into SWRL rules and directly implemented on the RDF knowledge graph through reasoner modules. The deduction of new and relevant information about each user and each vehicle item is then added to the dataset (see Figure \ref{fig_03}). The set of constraints relies on user preferences related to information about their favorite vehicles, which plays an essential role in finding relevant recommendations for the user. Therefore, we formulate these constraints using SPARQL queries based on pattern matching on graphs and solution modifiers, as shown in Figure \ref{fig_02}. In the ideal case, all the variables can be assigned, and we can find respondent solutions for the RDF knowledge graph. Therefore, we extract and recommend top-n vehicle recommendations from the results.
\begin{figure}[htbp]
	\centering
	\begin{tabular}{c c}
		\includegraphics[width=0.20\textwidth]{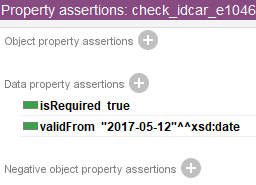} & \includegraphics[width=0.251\textwidth]{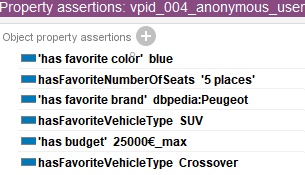}
	\end{tabular}
	
	\caption{Information deduced by using the SWRL rules translated from the constraints  $\mathcal{C}_{KB1}$ and $\mathcal{C}_{KB2}$.}
	\label{fig_03}
\end{figure}

\begin{figure}[htbp]
	
	\vspace{-0.5cm}
	\centerline{\includegraphics[width=0.35\textwidth]{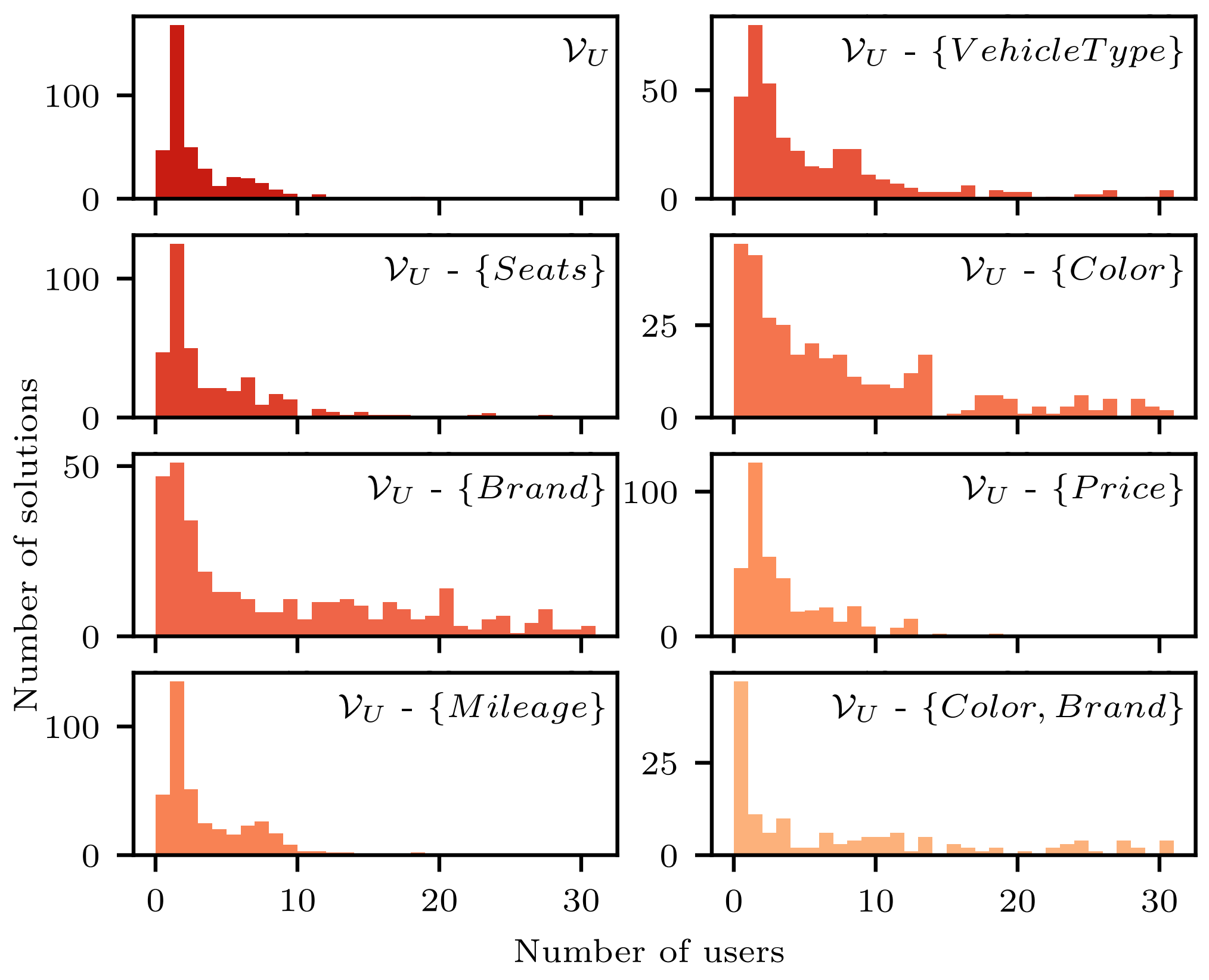}}
	\caption{Histograms representing the distribution of solutions across different sets of constraints.}
	\label{fig_04}
	\vspace{-0.3cm}
\end{figure}

However, many cases may not find a solution because of some inconsistencies from constraints between user preferences and vehicle descriptions. There are two possible propositions for this problem: (1)  Enriching and growing up the RDF knowledge graph by increasing the number of vehicles traded on portals; (2) Treating and identifying a minimal set of constraints from the user preference. Instead of enriching the dataset as in the first proposition, the second proposition relies on the elimination or adaptation of user constraints by using a diagnosis set which is defined as a constraint set $\Delta$ extracted from the set of constraints $\mathcal{C}_F$ such that the recommendations from the new set of constraints $\mathcal{C}_F$$-\Delta$ are consistent. 

To dive into our experiments, we build user preferences constraints based on the set of variables $\mathcal{V}_U$ $=$ $\{Seats,$ $VehicleType,$ $Brand,$ $Color,$   $Mileage,$ $Price\}$ extracted from the user preferences in the dataset. The experiment with all constraints based on user preferences resulted 88\% of users who found at least one solution from the RDF dataset.  With the second experiment, we aimed to build diagnosis sets in order to maximize the number of solutions matching user preferences and reduce the number of users who cannot find a solution. The diagnosis sets include only eliminated constraints from each user preference based on a preference order defined by the user on their preference. For instance, $\Delta_1 = \{Seats\}$, $\Delta_2 = \{VehicleType\}$, $\Delta_3 = \{Brand\}$, $\Delta_4 = \{Color\}$, $\Delta_5 = \{Mileage\}$, $\Delta_6 = \{Price\}$, $\Delta_7 = \{Color, Brand\}$. Figure \ref{fig_04} shows histograms about the distribution of the number of solutions over the number of users by using different diagnosis sets. With all constraints $\mathcal{V}_U$, the majority of the number of solutions is in a range from 0 to 5 solutions per user. By applying diagnosis sets, the number of solutions spreads out with an increase in the number of solutions greater than 10 for the users. These changes in the number of solutions are especially illustrated through the set of constraints: $\mathcal{V}_U - \Delta_3$ and $\mathcal{V}_U - \Delta_4$. In order to decrease the number of users who cannot find a solution, the elimination of several user preferences may become necessary. This means that we have to trade off user satisfaction with recommendation results from the recommender system.

We illustrate our constraint-based recommender system based on an RDF knowledge graph in the vehicle domain. The experiments demonstrate our proposal to separate the constraint sets into domain-specific knowledge constraint sets and user preference constraint sets. Using SWRL rules, the set of domain-specific knowledge can be inferred and integrated into the RDF dataset. The set of constraints built on user preference is translated into SPARQL queries. The relevant recommendations for users are extracted from solutions obtained through pattern matching on the RDF graphs.

\section{Conclusion and Perspectives}
In this paper, we investigate the construction of a constraint-based recommender system based on an RDF knowledge graph, which allows for the description and integration of information into a uniform model based on the support of ontologies. We illustrate how we can separate the constraint sets into domain-specific knowledge constraints and user preference constraints. Using SWRL rules, we translate domain-specific knowledge constraints into rules and make deductions of new relevant information on the RDF knowledge graph. User preference constraints can be directly translated into SPARQL queries. We conducted an experiment of our approach based on the RDF knowledge graph of the purchase and sales of vehicles. The recommendation results obtained from the constraint-based recommender system are promising. In our future work, we intend to research the exploitation of diagnosis sets , which should be optimized for each user and may help to reduce the performance time for proposing relevant recommendations. 



\section*{Acknowledgment}
This work was funded by the French Research Agency (ANR) and by the company Vivocaz under the project France Relance - preservation of R\&D employment (ANR-21-PRRD-0072-01).

\bibliographystyle{plain}
\bibliography{references} 

\end{document}